\documentstyle[aps,preprint,prl]{revtex}
\begin{document}
\flushbottom

\widetext
\draft
\title{Two Center Light Cone Calculation of Pair Production 
Induced by Ultrarelativistic Heavy Ions}
\author{A. J. Baltz}
\address{
Physics Department,
Brookhaven National Laboratory,
Upton, New York 11973}
\author{Larry McLerran}
\address{
Theoretical Physics Institute,
School of Physics and Astronomy,
University of Minnesota,
Minneapolis, Minnesota 55455}
\date{April 17, 1998}
%\date{\today}
\maketitle

\def\thepage{\arabic{page}}
\makeatletter
\global\@specialpagefalse
\ifnum\c@page=1
\def\@oddhead{Draft\hfill To be submitted to Phys. Rev. C}
\else
%\ifnum\c@page>1
\def\@oddhead{\hfill}
\fi
\let\@evenhead\@oddhead
\def\@oddfoot{\reset@font\rm\hfill \thepage \hfill}
\let\@evenfoot\@oddfoot
\makeatother

\begin{abstract}
An exact solution of the two center time-dependent Dirac equation for
pair production induced by ultrarelativistic heavy ion collisions is
presented.  Cross sections to specific final states approach those of
perturbation theory.  Multiplicity rates are reduced from perturbation theory.
\end{abstract}
%\newpage

\makeatletter
\global\@specialpagefalse
\def\@oddhead{\hfill}
\let\@evenhead\@oddhead
\makeatother
\nopagebreak
%\twocolumn
%\narrowtext

\section{Introduction}

In this paper, we shall compute the production of electron-positron pairs
in the central region for highly relativistic charged ions.  We base our
calculations on an exact solution of the time dependent Dirac equation in
the ultrarelativistic limit.  We show that except for possible cutoff effects
the exact cross section for any
specific final electron positron state equals the perturbation theory
result.  On the other hand, we argue that the rate for processes
correlated with low impact parameters (such as pair multiplicity) are reduced
from the perturbation theory rate.
 
Our notation
will vary at times between that of high energy physics (light cone variables
and Dirac $\gamma$ matrices) and that of atomic physic (Dirac 
$\mbox{\boldmath $\alpha$}$ and
$\beta$ matricies and the usual four momentum and space time variables) as
convenience and the connection to previous work dictates.  The relation between
different notations should be clear.

To properly 
define what we mean by central region and highly relativistic, it is useful
to define light cone coordinates,
\begin{equation}
                p^\pm = {1 \over \sqrt{2}} (p_0 \pm  p_z),
\ \ \ \ x^\pm = {1 \over \sqrt{2}} (t \pm z)
\end{equation}
where the z direction is the beam direction.  In this coordinate system,
the invariant dot product of momenta is $p \cdot q = p^+q^- + p^- q^+
- p_T \cdot q_T$.  We will take the nucleus propagating along the
positive z axis to have a large
light cone momentum $P^+$ and that along the negative z axis to have
large light cone momenta $Q^-$.  By highly relativistic we mean, for example,
the case of RHIC, colliding ions each with $\gamma\ (=1/\sqrt{1-v^2})$ of 100.
Note that the Compton wavelength of the electron is large compared to the
%Lorentz contracted 
radius of either nucleus and thus the nucleus can be considered a point charge
$\alpha Z$.
%\begin{equation}        
%             {1 \over {m_{electron}}} >> R/\gamma.
%\end{equation}

To define what we mean by centrally produced, we define the lightcone
momentum fractions of an electron or positron to be
\begin{equation}
                x = p^+ / P^+
\end{equation} 
and 
\begin{equation}
                y = p^-/Q^-
\end{equation}
Central production will mean electrons and positrons which have
$x, y << 1$  This can be satisfied for some range of longitudinal momenta
of the electron or positron so long as the condition on the 
Compton wavelength is satisfied.

When the electron or positron in the central region sees the moving
nuclei, it sees two oppositely Lorentz boosted
Coulomb fields.  These are the Li\'enard-Wiechert potentials
\begin{equation}
V(\mbox{\boldmath $ \rho$},z,t)={\alpha Z(1-v\alpha_z)\over
\sqrt{ [({\bf b}/2-\mbox{\boldmath $ \rho$})/\gamma]^2+(z-v t)^2}}
+{\alpha Z(1+v\alpha_z)\over
\sqrt{ [({\bf b}/2+\mbox{\boldmath $ \rho$})/\gamma]^2+(z+v t)^2}}.
\end{equation}
${\bf b}$ is the impact parameter, perpendicular to the $z$--axis along which
the ions travel, $\mbox{\boldmath $\rho$}$, $z$, and $t$ are the coordinates of
the potential relative to a fixed target (or ion)
$\alpha_z$ is the Dirac matrix,
and $Z, v$ and $\gamma$ are the charge, velocity and $\gamma$ factor of the
 oppositely moving ions.  We have specialized to the
case of equal $Z$ ions; unequal $Z$ would only require a trivial change in
what follows.  
%It has previously been
%shown that in the large $\gamma$ limit
%a gauge transformation on this potential can remove both the large
%positive and negative time contributions of the potential as well as the
%$\gamma$ dependence \cite{brw}.  
If one makes the gauge transformation on the wave function 
\begin{equation}
\psi=e^{-i\chi({\bf r},t)} \psi'
\end{equation}
where
\begin{equation}
\chi({\bf r},t)={\alpha Z \over v} 
\ln [\gamma(z-v t)+\sqrt{b^2+\gamma^2(z-v t)^2}]
-{\alpha Z \over v} 
\ln [\gamma(z+v t)+\sqrt{b^2+\gamma^2(z+v t)^2}]
\end{equation}
the interaction $V({\bf\rho},z,t)$ is gauge transformed to\cite{brw}
\begin{eqnarray}
V(\mbox{\boldmath $ \rho$},z,t)&=&{\alpha Z(1-v\alpha_z)\over
\sqrt{ [({\bf b}-\mbox{\boldmath $ \rho$})/\gamma]^2+(z-v t)^2}}
-{\alpha Z(1-(1/v)\alpha_z)\over\sqrt{b^2/\gamma^2+(z-v t)^2}}\nonumber \\
&&+{\alpha Z(1+v\alpha_z)\over
\sqrt{ [({\bf b}+\mbox{\boldmath $ \rho$})/\gamma]^2+(z+v t)^2}}
-{\alpha Z(1+(1/v)\alpha_z)\over\sqrt{b^2/\gamma^2+(z+v t)^2}}.
\end{eqnarray}
This gauge transformation reduces the range in $(z \pm v t)$ to more
closely map the ${\bf B}$ and ${\bf E}$ fields (which have the denomenator to
the ${3 \over 2}$ power rather than the ${1 \over 2}$ power of the
untransformed Lorentz gauge).  In the ultrarelativistic limit (ignoring
correction terms in $[({\bf b}+\mbox{\boldmath $ \rho$})/\gamma]^2$)
Eq.(7) takes the form\cite{ajb}
\begin{equation}
V(\mbox{\boldmath $ \rho$},z,t)
=\delta(z - t) (1-\alpha_z) \Lambda^-(\mbox{\boldmath $ \rho$})
+\delta(z + t) (1+\alpha_z) \Lambda^+(\mbox{\boldmath $ \rho$}) 
\end{equation}
where
\begin{equation}
\Lambda^{\pm}(\mbox{\boldmath $ \rho$}) = - Z \alpha 
\ln {(\mbox{\boldmath $ \rho$} \pm {\bf b}/2)^2 \over (b/2)^2}, 
\end{equation}
The potential as written here will be referred to the 
singular gauge solution.
In this gauge the field vanishes everywhere except along the lightcone
$x^\pm = 0$.  We can gauge transform from this field to the less singular
light cone gauge by again utilizing $\psi=e^{-i\chi({\bf r},t)} \psi'$
%\begin{equation}
%\psi=e^{-i\chi({\bf r},t)} \psi'
%\end{equation}
where
\begin{equation}
\chi({\bf r},t)
=\theta(t - z) \Lambda^-(\mbox{\boldmath $ \rho$})
+\theta(t + z) \Lambda^+(\mbox{\boldmath $ \rho$}) .
\end{equation}
This leads to added gauge terms in the transformed potential
\begin{eqnarray}
- {\partial \chi({\bf r},t) \over \partial t }
- \mbox{\boldmath $\alpha$} \cdot {\bf \nabla}
&=&-\delta(z - t) (1-\alpha_z) \Lambda^-(\mbox{\boldmath $ \rho$})
-\delta(z + t) (1+\alpha_z) \Lambda^+(\mbox{\boldmath $ \rho$})\nonumber \\ 
&&-\theta(t - z) \ \mbox{\boldmath $\alpha$} \cdot {\bf \nabla} 
 \Lambda^-(\mbox{\boldmath $ \rho$})
-\theta(t + z) \ \mbox{\boldmath $\alpha$} \cdot {\bf \nabla} 
\Lambda^+(\mbox{\boldmath $ \rho$}) 
\end{eqnarray}
and we thus obtain the light cone gauge
%This gauge transformation is
%A transformation can be made from this singular gauge to the light cone gauge
\begin{equation}
V(\mbox{\boldmath $ \rho$},z,t)
=-\theta(t - z) \ \mbox{\boldmath $\alpha$} \cdot {\bf \nabla} 
 \Lambda^-(\mbox{\boldmath $ \rho$})
-\theta(t + z) \ \mbox{\boldmath $\alpha$} \cdot {\bf \nabla} 
\Lambda^+(\mbox{\boldmath $ \rho$}) 
\end{equation}
with $\mbox{\boldmath $\alpha$}$ the Dirac matrix.  This construction fits in
well with the similar non-abelian treatment for QCD\cite{mv}.

%Defining light cone coordinates $x^\pm$ and fields $A^\pm$ in the obvious 
%way, these fields may be written as
%\begin{equation}
%        A^\pm = \delta(x^\mp) \Lambda^\mp (r_T)
%\end{equation}
%Here the vector potential contains both of the $\pm$ components above.
%The field $A^\pm$ arises from the nuclei with momentum along the
%$\pm$ $z$ direction. 

%We shall be interested in pairs which have transverse momenta
%much larger than the typical inverse size of the nuclei.  In this
%case, the nuclei appear as point charges as far as the transverse
%resolution is concerned.  If we let the impact parameter of the collision
%be $b$, then
%\begin{equation}
%        \Lambda^\pm (r_T) = Z^\pm \alpha ~ln\left(
%{{(\vec{r}_T - \vec{b}/2)^2} \over {(\vec{b}/2)^2}} \right)
%\end{equation}
%We will later specialize to the case of equal $Z$ ions.
%
%\begin{equation}
%        A^{\prime \mu}(x) = A^\mu (x) + \partial^\mu \Lambda (x)
%\end{equation}
%where
%\begin{equation}
%        \Lambda (x) = \theta (x^+) \Lambda^+ (x_T) + \theta (x^-)
%\Lambda^- (x_T)
%\end{equation}
%(Note that with our metric $\partial^\pm = \partial / \partial_\pm = 
%-\partial /\partial^\mp$.)
%The light cone gauge potential is therefore
%\begin{equation}
%        A^i = \theta(x^+)\nabla^i \Lambda^+(x_T) + \theta(x^-) \nabla^i
%\Lambda^- (x_T)
%\end{equation}
%Here the plus and minus components of the gauge potential vanish.

The reason why we expect that we can exactly compute pair production in the 
central region should be obvious.  In either of the gauges above,
the propagation of the electron or positron is essentially trivial.
Except at $x^\pm = 0$, the electron or positron propagates as a free
particle.  Therefore, to construct the propagator for the electron,
one needs to solve a boundary value problem with free propagation
everywhere except at the surfaces of discontinuity at the lightcone
$x^\pm = 0$.  

In fact to solve the problem of pairs production is arguably a little
simpler than constructing the propagator. What we will do is assume we begin
with a negative energy solution of the Dirac equation in the initial state.
This will correspond to a positive energy positron.
We will then let it propagate forward in time.  At late times we
will compute the amplitude that this state is a positive energy electron.

The organization of this paper is as follows:  In the second section,
we compute the amplitude for electron positron pair production in the singular
($\delta$ function) gauge.  The technique used will closely follow that
previously used in the
exact calculation of bound electron positron pair production\cite{ajbl},
and will use conventional Green's functions methods.  In the third section
we work in light cone gauge.  We show that the result for the pair production
amplitude agrees between the two computations.  In the fourth section,
we discuss the cross sections and their relation to perturbation theory.
In Appendix A, we write down our conventions for light cone coordinates and
projection operators in the Dirac equation.  In Appendix B, we evaluate the
transverse integral of the Coulomb field which is necessary for the solution
in both gauges.  In Appendix C, 
we show how to solve the Dirac equation across a boundary corresponding to a 
charged nucleus in light cone gauge.  We show how this result maps into the
similar result computed in singular gauge.  
  
\section{Solution in Singular Gauge for Pair Production}

A strategy to find the exact semi-classical solution for pair production in a
two-center model is to first find the exact wave function at the point of
interaction (in terms
of the appropriate Green's function) and then to construct the exact amplitude
(incorporating the initial exact wave function,
the interaction, and the final plane wave function).

In particular, one would like to solve the two center time-dependent Dirac
equation,
\begin{equation}
\Bigr[\mbox{\boldmath $\alpha$}{\bf p} + \beta - V(\mbox{\boldmath
 $\rho$},z,t) - E \Bigl]\Psi({\bf r},t) = 0,
\end{equation}
in terms of a Dirac Green's function based on the plane wave solutions,
\begin{equation}
\Bigr[\mbox{\boldmath $\alpha$}{\bf p} + \beta 
- E \Bigl]\phi({\bf r},t) = 0.
\end{equation}
One obtains
\begin{equation}
\Bigr[\mbox{\boldmath $\alpha$}{\bf p} + \beta 
- E \Bigl]\Psi({\bf r},t)
 = V(\mbox{\boldmath $\rho$},z,t) \Psi({\bf r},t)
+ \Bigr[\mbox{\boldmath $\alpha$}{\bf p} + \beta 
- E \Bigl]\phi({\bf r},t)
\end{equation}
If one acts on both sides with the Dirac plane wave Green's function
\begin{equation}
G_0 = \Bigr[\mbox{\boldmath $\alpha$}{\bf p} + \beta 
- E \Bigl]^{-1}
\end{equation}
one obtains the usual form
\begin{equation}
\Psi = \phi + G_0 V \Psi.
\end{equation}
The Green's function $G_0$ obeys the equation
\begin{eqnarray}
\Bigr[\mbox{\boldmath $\alpha$}{\bf p} + \beta 
- E \Bigl] G_0({\bf r},t;{\bf r^{\prime}},t^{\prime})&=&
\delta({\bf r - r^{\prime}}) \delta( t - t^{\prime})\nonumber \\
&=&{1 \over (2 \pi)^4} \int d^3 k  d \omega e^{i k ({\bf r - r^{\prime}})
-i \omega (t - t^{\prime})},
\end{eqnarray}
and the solution is
\begin{equation}
G_0({\bf r},t;{\bf r^{\prime}},t^{\prime}) = {1 \over (2 \pi)^4}
\int d^3 k  d \omega 
\Bigr[\mbox{\boldmath $\alpha$}{\bf k} + \beta + \omega \Bigl] 
{e^{i k ({\bf r - r^{\prime}})
-i \omega (t - t^{\prime})} \over k^2 +1 - \omega^2}.
\end{equation}
Note the positive sign of $\omega$ in the brackets.
We will get the boundary conditions later.

We now make use of the usual light cone coordinates, $x^+, x^-, q^+$, and $ q^-$, as in Eq.(1)
in place of $t, z, q_0$,and $q_z$.
Begin with a plane wave $\phi$ representing one of the electrons in the
filled negative energy Dirac sea.  Our convention will be that $q_0$ is
positive, corresponding to the positron energy we eventually are interested in,
and likewise the three momentum is opposite in sign to that of the negative
electron
\begin{equation}
\phi = v(q,s_i) e^{- i q x}= v(q,s_i) e^{-i q_{\perp}x_{\perp} + i q^+x^- + i
q^-x^+ }.
\end{equation}
The function $v(q,s_i)$ is the usual spinor of a positron.
The Green's function can likewise be written in light cone coordinates
\begin{equation}
G_0(x; x^{\prime}) = {1 \over (2 \pi)^4}
\int d k^+ dk^-  d^2 k_{\perp} 
\Bigr[\mbox{\boldmath $\alpha$}{\bf k} + \beta + \omega \Bigl] 
{e^{i k_{\perp} 
(x_{\perp} - x_{\perp}^{\prime}) -i k^+ (x^- - x^{\prime -})
-i k^- (x^+ - x^{\prime +}) }
\over - 2 k^+ k^- + k_{\perp}^2 +1 },
\end{equation}
and in light cone coordinates the two center time-dependent potential Eq.(8)
takes the form
\begin{equation}
V(x)= {1 \over \sqrt{2}} \delta(x^-) (1-\alpha_z) \Lambda^-(x_{\perp})
+ {1 \over \sqrt{2}}\delta(x^+) (1+\alpha_z) \Lambda^+(x_{\perp}).
\end{equation}
A space-time diagram is presented in Fig. 1.  Obviously $V(x)$ only acts on the
boundaries between the Regions I, II, III, and IV.

In Region I we have the initial plane wave $\phi$.  We would like to begin
by constructing the solutions in Regions II and III.  We will work out
Region III explicitly and then Region II will follow by symmetry.
We have from Eq. (17)
\begin{equation}
\Psi_{III}(x) = \phi(x) + \int d^4 z G_0(x,z) V(z) \Psi(z).
\end{equation}
But we have on the boundary between I and III\cite{ajbl}
\begin{equation}
V(z) \Psi(z) 
= {1 \over \sqrt{2}} \delta(z^-) (1-\alpha_z)
\Lambda^-(z_{\perp}) e^{i \theta (z^-) \Lambda^-(z_{\perp})} \phi(z), 
\end{equation}
which can be expressed equivalently
\begin{equation}
V(z) \Psi(z) 
=  {-i \over \sqrt{2}} \delta(z^-) (1-\alpha_z)
( e^{i \Lambda^-(z_{\perp})} - 1 ) \phi(z). 
\end{equation}
We now obtain
\begin{eqnarray}
\Psi_{III}(x)&=& \phi(x) - \int d z^+ d z^-  d^2 z_{\perp}
{ 1 \over (2 \pi)^4}
\int d k^+ d k^-  d^2 k_{\perp}  \nonumber \\
&\ & \times 
\Bigr[\alpha_{\perp} k_{\perp} + \beta 
+{1 \over \sqrt{2}}(1-\alpha_z) k^-
+{1 \over \sqrt{2}}(1+\alpha_z) k^+ \Bigl] {e^{i k_{\perp} 
(x_{\perp} - z_{\perp}) -i k^+ (x^- - z^-)  -i k^- (x^+ - z^+)}
\over - 2 k^+ k^- + k_{\perp}^2 +1 }
 \nonumber \\
&\ & \times {i \over \sqrt{2}} \delta(z^-) (1-\alpha_z)
( e^{i \Lambda^-(z_{\perp})} - 1 )
v(q,s_i) e^{-i q_{\perp}z_{\perp} + i q^+z^- + i q^-z^+ }.
\end{eqnarray}
The term in $(1+\alpha_z)$ obviously vanishes.  Now integrate over $z^-, z^+$,
and $k^-$ to obtain
\begin{eqnarray}
\Psi_{III}(x)&=& \phi(x) - \int { d k^+ d^2 k_{\perp} \over (2 \pi)^3}
\Bigr[\alpha_{\perp} k_{\perp} + \beta 
-{1 \over \sqrt{2}}(1-\alpha_z) q^- \Bigl]
{e^{i k_{\perp} x_{\perp}  -i k^+ x^-  +i q^- x^+ }
\over 2 k^+ q^- + k_{\perp}^2 +1 }
 \nonumber \\
&\ & \times {i \over \sqrt{2}} (1-\alpha_z)
\int d^2 z_{\perp} e^{-i (q_{\perp} + k_{\perp} )z_{\perp} }
( e^{i \Lambda^-(z_{\perp})} - 1 ) v(q,s_i). 
\end{eqnarray}

Now consider the integration over $k^+$.  We did not previously specify
the boundary condition on $G_0$ and we must do so now.  We want the $G_0$
term in Eq.(27) to be non-vanishing for $x^- > 0$ and to vanish for $x^- < 0$.
If we move the singularity just below the real axis by adding $ i \epsilon $
to the denominator then this physical boundary condition will be fulfilled.
We have 
\begin{equation}
\int_{-\infty}^{\infty}{ d k^+ e^{-i k^+ x^-} \over
k^+ + {k_{\perp}^2 +1 \over 2 q^-} +i \epsilon} 
= - 2 \pi i e^{i[(k_{\perp}^2 +1)/( 2 q^-)] x^-}
\end{equation}
and our expression for the wave function in Region III now becomes
\begin{eqnarray}
\Psi_{III}(x)&=& - \int { d^2 k_{\perp} \over (2 \pi)^2}
\Bigr[{1 \over \sqrt{2}}(\alpha_{\perp} k_{\perp} + \beta)(1-\alpha_z) 
- (1-\alpha_z) q^- \Bigl] 
 \nonumber \\
&\ & \times  
{e^{i k_{\perp} x_{\perp}  +i [(k_{\perp}^2 +1)/( 2 q^-)] x^-  +i q^- x^+ }
\over 2 q^-}
\int d^2 z_{\perp} e^{-i (q_{\perp} + k_{\perp} )z_{\perp} }
e^{i \Lambda^-(z_{\perp})} v(q,s_i). 
\end{eqnarray}
As a check on the validity of our expression we note that if we project with
$(1-\alpha_z)$ the expected boundary condition holds for
$x^- = + \epsilon$: 
\begin{equation}
(1-\alpha_z) \Psi_{III} = (1-\alpha_z) e^{i \Lambda^-(z_{\perp})} \phi.
\end{equation}

Now the exact amplitude takes the form
\begin{equation}
M = i \int dt < \phi \ | V \ | \Psi > .
\end{equation}
Begin by constructing $V$ times the exact wave function on the
boundary between Regions III and IV.  In analogy to Eq. (25) we have
\begin{equation}
V(x) \Psi(x) 
=  {-i \over \sqrt{2}} \delta(x^+) (1+\alpha_z)
( e^{i \Lambda^+(x_{\perp})} - 1 ) \Psi_{III}(x). 
\end{equation}
Note that we need the projection $(1+\alpha_z)$ for this boundary due to the
opposite direction of motion of the ion producing the $V$ as compared to the
$V$ on the boundary between I and III.
Thus we have at $x^+ = 0, x^- > 0$
\begin{eqnarray}
V(x) \Psi(x)&=& {i \over \sqrt{2}} \delta(x^+) (1+\alpha_z)
e^{i \Lambda^+(x_{\perp})} \int { d^2 k_{\perp} \over (2 \pi)^2}
{1 \over \sqrt{2}}(\alpha_{\perp} k_{\perp} + \beta)(1-\alpha_z) 
 \nonumber \\
&\ & \times  
{e^{i k_{\perp} x_{\perp}  +i [(k_{\perp}^2 +1)/( 2 q^-)] x^-  +i q^- x^+ }
\over 2 q^-}
\int d^2 z_{\perp} e^{-i (q_{\perp} + k_{\perp} )z_{\perp} }
e^{i \Lambda^-(z_{\perp})}  v(q,s_i). 
\end{eqnarray}
Likewise at $x^- = 0, x^+ > 0$
\begin{eqnarray}
V(x) \Psi(x)&=& {i \over \sqrt{2}} \delta(x^-) (1-\alpha_z)
e^{i \Lambda^-(x_{\perp})} \int { d^2 k_{\perp} \over (2 \pi)^2}
{1 \over \sqrt{2}}(\alpha_{\perp} k_{\perp} + \beta)(1+\alpha_z) 
 \nonumber \\
&\ & \times  
{e^{i k_{\perp} x_{\perp}  +i [(k_{\perp}^2 +1)/( 2 q^+)] x^+  +i q^+ x^- }
\over 2 q^+}
\int d^2 z_{\perp} e^{-i (q_{\perp} + k_{\perp} )z_{\perp} }
e^{i \Lambda^+(z_{\perp})}  v(q,s_i). 
\end{eqnarray}

In constructing the transition amplitude one makes use of the fact that only
two interaction (two photon) terms have a net contribution.  (Single
interaction terms integrated over the four boundaries give a null
contribution).
The amplitude then has two pieces
corresponding to the boundary of Region IV with Regions II and III
\begin{equation}
M(p,q) = M(p,q)_{IV,III} + M(p,q)_{IV,II}.
\end{equation}
The final state is a positive energy electron
\begin{equation}
\phi = u(p,s_f) e^{ i p x}= u(p,s_f) e^{i p_{\perp}x_{\perp} - i p^+x^- - i
p^-x^+ }.
\end{equation}
Then we have
\begin{eqnarray}
M(p,q)_{IV,III} &=& i \int_0^{\infty} d x^- \int d x^+ d x_{\perp}
\bar{u}(p,s_f) \beta
e^{- i p_{\perp}x_{\perp} + i p^+x^- + i p^-x^+ } 
 \nonumber \\
&\ & \times
{i \over \sqrt{2}} \delta(x^+) (1+\alpha_z)
( e^{i \Lambda^+(x_{\perp})} - 1 )\int { d^2 k_{\perp} \over (2 \pi)^2}
{1 \over \sqrt{2}}(\alpha_{\perp} k_{\perp} + \beta)(1-\alpha_z) 
 \nonumber \\
&\ & \times  
{e^{i k_{\perp} x_{\perp}  +i [(k_{\perp}^2 +1)/( 2 q^-)] x^-  +i q^- x^+ }
\over 2 q^-}
\int d^2 z_{\perp} e^{-i (q_{\perp} + k_{\perp} )z_{\perp} }
 e^{i \Lambda^-(z_{\perp})}  v(q,s_i).
\end{eqnarray}
Now integrate over $x^+$ and $x^-$ to obtain
\begin{eqnarray}
M(p,q)_{IV,III} &=& - \int { d^2 k_{\perp} \over (2 \pi)^2} 
{ \bar{u}(p,s_f) \beta (1+\alpha_z) (\alpha_{\perp} k_{\perp} + \beta) 
 v(q,s_i) \over  2 p^+ q^- + k_{\perp}^2 + 1}
 \nonumber \\
&\ & \times
\int d^2 x_{\perp} e^{ - i (p_{\perp} - k_{\perp}) x_{\perp} }
 e^{i \Lambda^+(x_{\perp})} 
\int d^2 z_{\perp} e^{-i (q_{\perp} + k_{\perp} )z_{\perp} }
 e^{i \Lambda^-(z_{\perp})} .
\end{eqnarray}
The transverse spatial integrals can be done in closed form as is shown in
Appendix B.
We obtain
\begin{equation}
\int d^2 y_{\perp} e^{-i k_{\perp} y_{\perp} }
 e^{i \Lambda^{\pm}(y_{\perp})} = e^{-i k_{\perp} y_{\perp} } 
\biggl({b^2 \over 16}\biggr)^{ i \eta}
{4 \pi \over k_{\perp}^{2 - 2 i \eta} }{\Gamma (1 - i \eta) \over 
\Gamma( i \eta)}
\end{equation}
where $\eta = +Z \alpha$.

The amplitude now becomes
\begin{eqnarray}
M(p,q)_{IV,III} &=& - 4 {\Gamma^2 (1 - i \eta) \over \Gamma^2 ( i \eta)}
\int  d^2 k_{\perp} 
{ \bar{u}(p,s_f) \beta (1+\alpha_z) (\alpha_{\perp} k_{\perp} + \beta) 
 v(q,s_i) \over  2 p^+ q^- + k_{\perp}^2 + 1}
 \nonumber \\
&\ & \times e^{ - i b (p_{\perp}/2 - q_{\perp}/2 -k_{\perp})}
%{1 \over (p_{\perp} - k_{\perp})^2}
%{1 \over (k_{\perp} + q_{\perp})^2}
%\biggl( {b^2 (p_{\perp} - k_{\perp})^2 \over 16}\biggr)^{i \eta}
%\biggl( {b^2 (k_{\perp} + q_{\perp})^2 \over 16}\biggr)^{i \eta}
\bigl( (p_{\perp} - k_{\perp})^2 (k_{\perp} + q_{\perp})^2 \bigr)^{i \eta - 1}
\end{eqnarray}
where trivial $b$ dependent and constant phases have been removed.

The other piece of the amplitude coming from the boundary between Region II
and Region IV has the corresponding form
\begin{eqnarray}
M(p,q)_{IV,II} &=& - 4 {\Gamma^2 (1 - i \eta) \over \Gamma^2 ( i \eta)}
\int d^2 k_{\perp} 
{ \bar{u}(p,s_f) \beta (1-\alpha_z) (\alpha_{\perp} k_{\perp} + \beta) 
 v(q,s_i) \over  2 p^- q^+ + k_{\perp}^2 + 1}
 \nonumber \\
&\ & \times e^{ + i b (p_{\perp}/2 - q_{\perp}/2 -k_{\perp})}
\bigl( (p_{\perp} - k_{\perp})^2 (k_{\perp} + q_{\perp})^2 \bigr)^{i \eta - 1}
 \nonumber \\
&=& - 4 {\Gamma^2 (1 - i \eta) \over \Gamma^2 ( i \eta)}
\int d^2 k_{\perp} 
{ \bar{u}(p,s_f) \beta (1-\alpha_z) (\alpha_{\perp} (p_{\perp} - q_{\perp}
 -k_{\perp}) + \beta) 
 v(q,s_i) \over  2 p^- q^+ +
 (p_{\perp} - q_{\perp} -k_{\perp})^2 + 1}
 \nonumber \\
&\ & \times e^{ - i b (p_{\perp}/2 - q_{\perp}/2 -k_{\perp})}
\bigl( (p_{\perp} - k_{\perp})^2 (k_{\perp} + q_{\perp})^2 \bigr)^{i \eta - 1},
\end{eqnarray}
and the total amplitude takes the form
\begin{eqnarray}
M(p,q)&=& 4 \eta^2
\int d^2 k_{\perp} \ e^{ i b k_{\perp}}\  
\Bigl( (p_{\perp} - k_{\perp})^2 (k_{\perp} + q_{\perp})^2 \Bigr)^{i \eta - 1}
\nonumber \\
& \times & \biggl({ \bar{u}(p,s_f) \beta (1+\alpha_z) (\alpha_{\perp} k_{\perp}
 + \beta)  v(q,s_i) \over  2 p^+ q^- + k_{\perp}^2 + 1}
\nonumber \\
&& +\  { \bar{u}(p,s_f) \beta (1-\alpha_z) (\alpha_{\perp} (p_{\perp} 
- q_{\perp}
 -k_{\perp}) + \beta) 
 v(q,s_i) \over  2 p^- q^+ +
 (p_{\perp} - q_{\perp} -k_{\perp})^2 + 1} \biggr) 
\end{eqnarray}
where trivial phases depending on $\eta$ and on initial and final momenta
have been removed.

Rewriting with the mass of the electron explicit, and making use of more
modern notation we have
\begin{eqnarray}
M(p,q)&=& 4 \eta^2
\int d^2 k_{\perp} \ e^{ i b k_{\perp}}\  
\Bigl( (p_{\perp} - k_{\perp})^2 (k_{\perp} + q_{\perp})^2 \Bigr)^{i \eta - 1}
\nonumber \\
& \times & \biggl({ \bar{u}(p,s_f) (1-\alpha_z) ( - \not\!k_{\perp}
 + m)  v(q,s_i) \over  2 p^+ q^- + k_{\perp}^2 + m^2}
\nonumber \\
&& +\  { \bar{u}(p,s_f) (1+\alpha_z)  ( - \not\!p_{\perp} +
 \not\!q_{\perp} + \not\!k_{\perp} + m) 
 v(q,s_i) \over  2 p^- q^+ +
 (p_{\perp} - q_{\perp} -k_{\perp})^2 + m^2} \biggr). 
\end{eqnarray}

At this point one notices that the infinite range of the transverse potential
provides us with a result lacking an infrared cutoff.  
In Appendix B, we introduce a convergence parameter $\epsilon$ to regularize
this infrared singular behavior.
Strictly speaking we
were not allowed to let the convergence parameter $\epsilon$ take on the value
zero.  In the previous use of
the singular interaction for calculating bound electron positron pair
production, the transverse integrals were cut off by the finite range of the
bound state wave function\cite{ajbl}.  With plane waves there is no such
natural cutoff.  This suggests that we attach a physical interpretation to
the convergence factor $\epsilon$ in Appendix B.  In fact if we set $\epsilon$
equal to $\omega / \gamma$, where $\omega$ is the energy of the produced
electron or positron, then we obtain the expected spatial cutoff in a heavy ion
electromagnetic interaction, and a
result that can naturally be compared with the corresponding perturbation
theory result of Bottcher and Strayer\cite{stra}.  We discuss this further in
Section IV.  But first we shall come to the same solution in the light
cone gauge.

\section{The Solution to the Two Charge Problem in Light Cone Gauge}

In this section, we present an alternative derivation to the results presented
above in singular gauge.  This serves as a check on the results of
the previous section.  Also the technique which we present here
appears to generalize directly to the non-abelian problem.

We begin by considering a negative energy plane wave.  Here we will take
$q^\pm <0$.  This negative energy incoming plane wave will eventually be
interpreted as an incoming positron with momentum $q \rightarrow -q$.
So we begin with an incoming wave
\begin{equation}
	\psi_I (x) = e^{iqx} v_\lambda (-q).
\end{equation} 
Here $v_\lambda$ is a positron spinor.  The label $I$ on the solution refers
to the fact that this is a solution in region $I$ as in Fig. 1, and as in
Section II.

Recall that the background field in light cone gauge is of the
form
\begin{equation}
	A^i(x) = -\theta (x^-) \nabla^i \Lambda^- (x_T)
  - \theta (x^+) \nabla^i \Lambda^+ (x_T)
\end{equation}
as was discussed in the introduction.  

Let us first construct the solution in region $II$.  This
involves constructing the solution in region across the boundary at $x^+ = 0$
from region $I$.  Everywhere in region $II$, $x^- < 0$.  
Using the results of 
Appendix C, a general solution is of the form
\begin{eqnarray}
	\psi_{II} (x) &  = & {1 \over \sqrt{2}}
e^{-i\Lambda^+ (x_T)} \int {{dp^+ d^2p_T}
\over {(2\pi)^3}}  
e^{\{-ip^+x^-+ip_Tx_T - i (p_T^2+m^2)x^+/2(p^++i\epsilon)\}} 
\left\{1 + {{\alpha_T \cdot p_T +m} \over {\sqrt{2}(p^++i\epsilon)}} \right\}
\nonumber \\
 & & \alpha^+ v_\lambda (-q) 
\left\{ {1 \over i} {1 \over {p^+-q^+-i\epsilon}}
\int d^2z_T e^{i(q_t-p_T)z_T} e^{i\Lambda^+ (z_T)}
+ H(p_T, p^+, q) \right\} 
\end{eqnarray}
In this equation, H has all its singularities in the negative half
$p^+$ plane.  Therefore at $x^+ = 0$, this contribution gives nothing.
In fact throughout region $II$, there is no contribution so that 
$H$ can be dropped.
The choice of singularity for $p^+$ in the exponential is
so that there is convergence at large positive $x^+$.  Notice that for 
$x^- < 0$ corresponding to region $II$, 
the term which does not involve $H$ may be 
evaluated by closing in the upper half $p^+$ plane.  Here the solution
satisfies the correct boundary condition at $x^+ = 0$ and solves the 
Dirac equation in region $II$.  It is unique.

In exactly the same way, we have in region $III$, that the solution is
\begin{eqnarray}   
\psi_{III} (x) &  = & {1 \over \sqrt{2}}
e^{-i\Lambda^- (x_T)} \int {{dp^- d^2p_T}
\over {(2\pi)^3}}  
e^{\{-ip^-x^++ip_Tx_T - i (p_T^2+m^2)x^-/2(p^-+i\epsilon)\}} 
\left\{1 + {{\alpha_T \cdot p_T +m} \over {\sqrt{2}(p^-+i\epsilon)}} \right\}
\nonumber \\
 & & \alpha^- v_\lambda (-q) 
\left\{ {1 \over i} {1 \over {p^--q^--i\epsilon}}
\int d^2z_T e^{i(q_t-p_T)z_T} e^{i\Lambda^- (z_T)}
\right\} 
\end{eqnarray}

We now write down an ansatz for the solution in region $IV$.
After writing down the solution, we will show that it solves the Dirac
equation and the boundary conditions at $x^\pm = 0$ when $x^\mp > 0$.
Consider
\begin{eqnarray}
	\psi_{IV} (x) & = & {i \over 2} e^{-i\Lambda^+(x_T)-i\Lambda^-(x_T)}
\int {{d^2p_T d^2z_T} \over {(2\pi)^2}} {{d^2p^\prime_Td^2z^\prime_T}
\over {(2\pi)^2}} \left\{ \right. \nonumber \\ 
& & \int {{dp^+} \over {2\pi}} {1 \over {p^+ - (p_T^2+m^2)/2q^- +
i\epsilon}} \nonumber \\ 
& & e^{-ip^+x^- +ip^\prime_Tx_T 
-i(p^{\prime 2}_T + m^2)x^+/2(p^+ +i\epsilon)}  \nonumber \\
& & e^{iz^\prime_T(p_T - p^\prime_T) + i z_T (q_T - p_T)} e^{i\Lambda^+(z_T)
+i\Lambda^-(z^\prime_T)} \nonumber \\
& & \left(1 + {{\alpha_T \cdot p^\prime_T + \beta m} \over
{\sqrt{2} (p^++i\epsilon)}} \right) \alpha^+
{{\alpha_T \cdot p_T + \beta m } \over {\sqrt{2} q^-}} \alpha^- 
v_\lambda (-q) \nonumber \\
& & +  \int {{dp^-} \over {2\pi}} {1 \over {p^- - (p_T^2+m^2)/2q^+ +
i\epsilon}} e^{-ip^-x^+ +ip^\prime_Tx_T 
-i(p^{\prime 2}_T + m^2)x^+/2(p^- +i\epsilon)}  \nonumber \\
& & e^{iz^\prime_T(p_T - p^\prime_T) + i z_T (q_T - p_T)} e^{i\Lambda^-(z_T)
+i\Lambda^+(z^\prime_T)} \nonumber \\
& & \left(1 + {{\alpha_T \cdot p^\prime_T + \beta m} \over
{\sqrt{2} (p^++i\epsilon)}} \right) \alpha^-
{{\alpha_T \cdot p_T + \beta m } \over {\sqrt{2} q^+}} \alpha^+ 
v_\lambda (-q) \left. \right\}
\end{eqnarray}

First it is straightforward to apply the Dirac equation to $\psi_{IV}$,
and show that it solves the Dirac equation.  What about the behavior on
the boundaries $x^\pm = 0$, $x^\mp > 0$.  First consider $x^- = 0$ and
$x^+ > 0$.  We first evaluate $\psi_{II}$ on this boundary.  We require that
the $\psi^-$ be continuous.  Using $\psi_{II}$, we see that
\begin{eqnarray}
	\psi_{II}^-\mid_{x^- = 0, x^+ > 0} & = & {1 \over {\sqrt{2}}}
e^{-i\Lambda^+(x_T)} \int {{d^2p_T} \over {(2\pi)^2}}
e^{ip_Tx_T - i (p_T^2 + m^2)x^+/2q^+} \nonumber \\
& & {{\alpha_T \cdot p_T + \beta m} \over {\sqrt{2} q^+}}
\alpha^+ v_\lambda (-q) \int d^2z_T e^{i(q_T-p_T)z_T} e^{i\Lambda^+(z_T)}
\end{eqnarray}
 
Now, how does $\psi^-_{IV}$ behave on this boundary?  The first integral
over $p^+$ in the integral representation for $\psi^-_{IV}$ vanishes at
$x^- = 0$ as the all the singularities of the integration over $p^+$ are
in the same side of the  real $p^+$ axis in the complex $p^+$ plane.
In the second term which involves $p^-$, we must close in the lower half
$p^-$ planes when $x^- = 0$ and $x^+ > 0$.  Closing in this plane and doing
the integrals over $z_T^\prime$ and $p_T^\prime$ shows that our formula 
solves the correct boundary condition.

By symmetry, we see that along the other boundary at $x^+ = 0$ for $x^- > 0$
the boundary condition is solved for the plus component of the wavefunction.
Therefore $\psi_{IV}$ solves the Dirac equation and the boundary conditions
in region $IV$.

Now we must extract the matrix element for pair production from this 
wavefunction.  To do this, we observe that after a little algebra,
and making the substitution $q \rightarrow -q$ that
\begin{eqnarray}
	\psi_{IV} (x) & = & -{1 \over 2} e^{-i\Lambda^+(x_T) - i\Lambda^-(x_T)}
\int {{d^4p^\prime} \over {(2\pi)^4}} {{d^2p_T} \over {(2\pi)^2}}
d^2z_T d^2z^\prime_T \nonumber \\
& & {{m- p^\prime \cdot \gamma} \over {p^{\prime 2} + m^2 -i\epsilon}}
e^{-ip^\prime x} \left\{ \right. \nonumber \\
& & {{2p^{\prime +}} \over {p^{\prime +} + (p_T^2 + m^2)/2q^- +i\epsilon}}
e^{iz_T^\prime(p_T-p^\prime_T) -iz_T(q_T+p_T)} \nonumber \\
& & e^{i\Lambda^+(z^\prime_T) + i\Lambda^-(z_T)}
\alpha^- {{m+p_T \cdot \gamma} \over {\sqrt{2} q^-p^{\prime +}}} v_\lambda (q) 
\nonumber \\
& & {{2p^{\prime -}} \over {p^{\prime -} + (p_T^2 + m^2)/2q^+ +i\epsilon}}
e^{iz_T^\prime(p_T-p^\prime_T) -iz_T(q_T+p_T)} \nonumber \\
& & e^{i\Lambda^-(z^\prime_T) + i\Lambda^+(z_T)} 
\alpha^+ {{m+p_T \cdot \gamma} \over {\sqrt{2} q^+p^{\prime -}}} v_\lambda (q) 
\left. \right\} \\
\end{eqnarray}

Amputating the external line which corresponds to the propagator for
the electron, we find the matrix element for pair production to be
\begin{eqnarray}
M_{\eta \lambda} & = & \int {{d^2p^\prime_T} \over {(2\pi)^2}}
\left\{ \right. 
\kappa^+(p^\prime_T - p_T) \kappa^- (p_T^\prime+q_T)
{1 \over {2q^-p^{\prime +} + p^{\prime 2}_T + m^2 +i\epsilon}}
{\overline u}_\eta (p) \sqrt{2} \alpha^- ({m+p_T^\prime \cdot \gamma}  )
v_\lambda(q) \nonumber \\
& & + \kappa^+(p^\prime_T +q_T) \kappa^- (p_T^\prime -p_T)
{1 \over {2q^+p^{\prime -} + p^{\prime 2}_T + m^2 +i\epsilon}}
{\overline u}_\eta (p) \sqrt{2} \alpha^+ ({m+p_T^\prime \cdot \gamma}  )
v_\lambda(q) \left. \right\}
\end{eqnarray}
Here the quantity
\begin{equation}
	\kappa^\pm(p_T) = \int d^2x_T e^{-ip_Tx_T} e^{+i\Lambda^\pm (x_T)} .
\end{equation}

\section{Discussion}

After we had finished the derivation of the amplitudes Eq.(43) and its
equivalent Eq.(52) we became aware of similar results recently obtained by
Segev and Wells\cite{sw}.  These authors obtained an amplitude with the same
structure as ours, but they did not obtain our closed form, Eq. (40) (B6) or
Eq.(B5) for the non-perturbative transverse momentum integral.  They did
argue that the transverse integral went appropriately to the ultrarelativistic
perturbative limit in agreement with the result of Bottcher and
Strayer\cite{stra}.  In this limit their perturbative amplitude simply takes
the form of Eq.(43) without the $i \eta$ in the exponent.

We now have the suggestion of an ansatz for a physical cutoff of the exact
result in line with the discussion at the end of Section II.  The perturbation
theory result of Bottcher and Strayer not taken to the ultrarelativistic limit
actually has denominators of the form $k_\perp^2 + \omega^2 / (\gamma^2 -1)$
rather than $k_\perp^2$.  Our ansatz then is to modify Eq.(43) to
\begin{eqnarray}
M(p,q)&=& 4 \eta^2
\int d^2 k_{\perp} \ e^{ i b k_{\perp}}\  
\biggl(\Bigl( [p_{\perp} - k_{\perp}]^2 + {\omega^2 \over \gamma^2}\Bigr)
\Bigl( [k_{\perp} + q_{\perp}]^2 + {\omega^2 \over \gamma^2}\Bigr)
\biggr)^{i \eta - 1}\nonumber \\
& \times & \biggl({ \bar{u}(p,s_f) (1-\alpha_z) ( - \not\!k_{\perp}
 + m)  v(q,s_i) \over  2 p^+ q^- + k_{\perp}^2 + m^2}
\nonumber \\
&& +\  { \bar{u}(p,s_f) (1+\alpha_z)  ( - \not\!p_{\perp} +
 \not\!q_{\perp} + \not\!k_{\perp} + m) 
 v(q,s_i) \over  2 p^- q^+ +
 (p_{\perp} - q_{\perp} -k_{\perp})^2 + m^2} \biggr). 
\end{eqnarray}
We have effectively retained $\epsilon$ (set to $\omega / \gamma$) in the
factor taken to the $1 - i \eta$ power of Eq.(B5) and ignored it elsewhere.
Retaining $\eta$ elsewhere in Eq.(B5) would cut off a little more sharply at
small $k_\perp$ but is not necessary.  It would not necessarily be more exact
to retain all factors of $\eta$ because the cutoff comes in response to the
spatial region 
$\mbox{\boldmath $ \rho$}   = \gamma / \omega$ where both the singular and
light cone potentials begin to lose their validity.  

At this point one is left with a choice whether to perform the integral over
the impact parameter either before or after the integral over the intermediate
transverse momentum.  If one is only interested in a the cross section for
pair production in a given small momentum and/or energy bin then one must
obtain the same answer independent of the order of integration.  In that case
performing the integral over the impact parameter first seems convenient.
One has an impact parameter integral of the form
\begin{equation}
\int d^2 {\bf b} \int \vert M(p,q) \vert^2 \sim \int d^2 {\bf b}
\int d^2 k_{\perp} e^{ i b k_{\perp}} f(k_{\perp})
\int d^2 k_{\perp}^{\prime} e^{ - i b k_{\perp}^{\prime}} 
f^*(k_{\perp}^{\prime}).
\end{equation}
The integral over $d^2 {\bf b}$ gives $(2 \pi)^2 \delta^{(2)} (k_{\perp} -
k_{\perp}^{\prime})$ and the $\pm i \eta $ exponents vanish, giving a
result identical to what we would obtain in perturbation theory.

If we are interested in high multiplicity events or a total cross section for
events in which at least one pair is produced then we must perform the
integral over the intermediate transverse momentum first.  For the highest
multiplicity events we need
to evaluate the square of the amplitude at small but non-intersecting values
of the impact parameter: we need the mean number of pairs at a given impact
parameter.  From the mean number of pairs produced at each impact parameter,
one can obtain the probability (less than one) of at least one pair
being produced.  Thus a total pair production cross section can be defined
and computed.

Note that for a given impact parameter the square of the amplitude is not
identical in the exact solution to what it is in the the perturbation theory
solution.  For the exact solution the $i\ \eta$ in the exponent gives a
rotating phase in the the integral over $k_{\perp}$ that is absent in
the perturbative case.  For example if we look at a dominant contribution near
one of the cutoffs the amplitude goes as
\begin{equation}
M \sim \int_0^k { k_\perp d k_\perp \over (k_\perp^2 + \omega^2 / 
\gamma^2)^{1 - i \eta}} = {1 \over 2 i \eta}(e^{i \eta \ln(k^2 +
 \omega^2/\gamma^2 )} - e^{i \eta \ln{\omega^2/\gamma^2 }})
\end{equation}
which decreases with increasing $\eta$.  Here the exact contribution
is less than perturbation theory would be.

To sum up, we have a situation where one can say both that the exact cross
section for pair production to any final state is identical to the
perturbation theory cross section, and that measurements of high multiplicity
pair events will show deviations from perturbation theory probably by being
smaller.

\section{Acknowledgements}

We thank the Institute for Nuclear Theory at the University of Washington for
its hospitality during the Fall 1996 program, ``Ultrarelativistic Nuclei: From
Structure Functions to the Quark-Gluon Plasma.''  This work was initiated
during that program.
We would like to thank D. Kharzeev for useful discussions.  A. J. B. would
like to thank A. Gal for drawing our attention to the work of of Segev and
Wells.
\vskip .5cm
This manuscript has been authored under Contracts No. DE-AC02-98CH10886,
DE-AC02-83ER40105, and DE-FG02-87ER-40328 with
the U. S. Department of Energy. 
\appendix
\section{Conventions for Light Cone Variables and the Use of Projection
Operators}

We will let the light cone coordinates for any vector be
\begin{equation}
	V^\pm = {1 \over \sqrt{2}} (V^0 \pm V^z).
\end{equation}
The metric convention we will use is $g^{00} = 1$ and $g^{ij} = - \delta^{ij}$
for spatial components so that in light cone gauge the dot product is
\begin{equation}
	A \cdot B = A^+ B^- + A^- B^+ - A_T \cdot B_T.
\end{equation}
We shall use ordinary gamma matrices for fermions so that
\begin{equation}
	\{ \gamma^\mu, \gamma^\nu \} =  2 g^{\mu \nu}.
\end{equation}

We can define projection operators for fermions as
\begin{equation}
P^\pm = {1 \over \sqrt{2}} \gamma^0 \gamma^\pm = {1 \over 2}(1 \pm \alpha_z).
\end{equation}
These projection operators satisfy $P^+ + P^- =1$, $P^{\pm 2} = P^\pm$,
and $P^\pm P^\mp = 0$.  The plus and minus components of a fermion field are
\begin{equation}
	\psi^\pm = P^\pm \psi
\end{equation}
where $\psi^+ + \psi^- = \psi$.

In light cone coordinates, the Dirac equation for a free particle becomes
\begin{equation}
	\psi^- = {1 \over {\sqrt{2} p^+}} (p_T \cdot \alpha_T + \beta m) 
\psi^+
\end{equation}
where $\psi^+$ solve
\begin{equation}
	(2p^+p^- -p_T^2 -m^2)\psi^+ = 0.
\end{equation}
In an external field which is independent of $x^\pm$, 
the transverse momentum is
converted into a transverse covariant momentum.
Also, the relationships can be reversed under the transformation
$\psi^\pm \rightarrow \psi^\mp$.

\section{Evaluation of the Transverse Integral}
We would like to evaluate the integral
\begin{equation}
I = \int d^2  \rho
\ e^{-i {\bf k} \cdot \rho}
e^{-i \eta \ln {( \rho \pm {\bf b}/2)^2
 \over (b/2)^2}}.
\end{equation}
A coordinate shift gives us
\begin{equation}
I = \biggl({b^2 \over 4}\biggr)^{ i \eta}\ e^{\mp i [{\bf k} \cdot {\bf b} /2]}
\int_0^{2 \pi} d \phi \int_0^{\infty} \rho d \rho \ e^{-(i k_{\perp}
 \cos{\phi} + \epsilon |\cos{\phi}|) \rho } e^{-i \eta \ln{\rho}^2},
\end{equation}
where we have added a convergence factor $\epsilon |\cos{\phi}|$.
The integral over $\rho$ can then be carried out (\cite{gr}
3.381.5) and we obtain
\begin{equation}
I = \biggl({b^2 \over 4}\biggr)^{ i \eta}\ e^{\mp i [{\bf k} \cdot {\bf b} /2]}
\int_0^{2 \pi} d \phi {\Gamma (2 - 2 i \eta) \over[ (k_{\perp}^2
+ \epsilon^2 ) \cos^2{\phi} ]^{1 - i \eta}} 
e^{ (- 2 i - 2 \eta)\arctan{( k_{\perp} \cos{\phi} / \epsilon |\cos{\phi}|)}}.
\end{equation}
Making use of quadrant symmetry this may be rewritten
\begin{eqnarray}
I &=&4\ e^{\mp i[{\bf k}\cdot {\bf b}/2]} \biggl({b^2 \over 4}\biggr)^{ i \eta}
\ { (\epsilon^2 - k_{\perp}^2) 
\cosh(2 \eta \arctan{ [k_{\perp} / \epsilon}])
+ i 2 \epsilon k_{\perp} \sinh(2 \eta \arctan{ [k_{\perp} / \epsilon}])
\over (\epsilon^2 + k_{\perp}^2 ) } \nonumber \\
&&\times{\Gamma (2 - 2 i \eta) \over (k_{\perp}^2 + \epsilon^2)^{1 -  i \eta}} 
\int_0^{ \pi / 2} { d \phi \over (\cos{\phi})^{2 - 2 i \eta}}. 
\end{eqnarray}
The integral over $\phi$ can then be carried out by Wallis' Formula
(\cite{as} 6.1.49) giving
\begin{eqnarray}
I &=&4\ e^{\mp i[{\bf k}\cdot {\bf b}/2]} \biggl({b^2 \over 4}\biggr)^{ i \eta}
\ { (\epsilon^2 - k_{\perp}^2) 
\cosh(2 \eta \arctan{ [k_{\perp} / \epsilon}])
+ i 2 \epsilon k_{\perp} \sinh(2 \eta \arctan{ [k_{\perp} / \epsilon}])
\over (\epsilon^2 + k_{\perp}^2 ) } \nonumber \\
&&\times{\Gamma (2 - 2 i \eta) \over (k_{\perp}^2 + \epsilon^2)^{1 -  i \eta}} 
{\sqrt{\pi}\ \Gamma( - .5 + i \eta ) \over 2 \Gamma( i \eta)}.
\end{eqnarray}
Letting $\epsilon$ approach zero from the positive direction and exploiting
$\Gamma$ function relations finally leads to
\begin{equation}
I = - e^{\mp i[{\bf k}\cdot {\bf b}/2]} \biggl({b^2 \over 16}\biggr)^{ i \eta}
{4 \pi \over k_{\perp}^{2 - 2 i \eta} }{\Gamma (1 - i \eta) \over 
\Gamma( i \eta)}. 
\end{equation}

\section{Solving the Dirac Equation for the Li\'enard-Wiechert Potential}

In this appendix, we discuss solving the Dirac equation for Li\'enard-Wiechert
potentials appropriate for the central region.  We will show how to construct
a solution for a light cone potential of the form
\begin{equation}
	A^i_T = - \theta (x^-) \nabla^i \Lambda (x_T)
\end{equation}
We will then show how to convert this solution into that for
the singular gauge and how this problem in singular gauge translates
into a boundary condition on the Dirac wavefunction at $x^- = 0$.
In the text, we will use this to construct solutions across the boundaries at
$x^\pm = 0$.

We assume that for $x^- < 0$, we are given a plane wave solution of the form
\begin{equation}
	\psi(x) = e^{iqx} u_\lambda (x)
\end{equation}
where $\lambda$ is a polarization label on the electron spinor.
At $x^- > 0$, the solution must be a linear combination of plane wave
solutions to the Dirac equation modulo a gauge rotation
\begin{equation}
	\psi^\prime (x) = e^{i\Lambda (x_T)} \int
{{d^2p_T} \over {(2\pi)^2}}  F_{\lambda \lambda^\prime} (p_T, q_T)
u_\lambda^\prime (p) e^{(ip_T x_T -iq^-x^+ -ip^+x^-)}
\end{equation}
where
\begin{equation}
	p^+ = {{p_T^2+m^2} \over {2q^-}}.
\end{equation}
The integral above is two dimensional since $q^-$ labels the solution on both
sides of the boundary as the potential is $x^+$ independant, and because
the value of plus component of light cone momenta is determined by the
mass shell condition.

What are the boundary conditions at $x^- = 0$?   Recall that the form of the
Dirac equation is
\begin{equation}
	(P_T\cdot \alpha_T + \beta m - \sqrt{2} p^+ P^- - \sqrt{2} p^- P^+ )
\psi = 0
\end{equation}
where $P^\pm$ are the projection operators described in the last appendix.
Here $P_T$ is the covariant momentum operator.
We see there for the $\psi^-$ can be chosen to be continuous across the 
boundary.  On the other hand, the plus component of the wavefunction must be
discontinuous since by the Dirac equation
\begin{equation}
	\psi^+ = {1 \over {\sqrt{2} p^-}} (P_T \cdot \alpha_T + \beta m) 
\psi^-
\end{equation}
and the covariant momentum operator contains the vector potential which is
discontinuous.

This tells us what the proper boundary condition is in singular gauge.
Recall that to transform between the two gauges
\begin{equation}
	\psi_{singular} (x) = e^{i\theta(x^-)\lambda (x_T)} 
\psi_{lightcone} (x).
\end{equation}
We therefore have that the boundary condition in singular gauge is the
the component 
\begin{equation}
	\psi_{singular}^-(x) \mid_{x^- = 0^+} =
  e^{i\lambda(x_T)} \psi_{singular}^- (x) \mid_{x^- = 0^-}
\end{equation}
with the plus components determined by the Dirac equation.
All components of the wavefunction 
in singular gauge are therefore discontinuous.

It is now straightforward to determine the solution for $x^- > 0$.
Consider
\begin{eqnarray}
	\psi(x) & = & {1 \over \sqrt{2}}
e^{i\Lambda (x_T)} \int {{d^2p_T} \over {(2\pi)^2}}
d^2z_T e^{-i\Lambda(z_T)} e^{ip_Tx_T - i q^-x^+ -ip^+x^-}
e^{iz_T(q_T-p_T)} \nonumber \\
 & & \left\{1 + {{\alpha_T \cdot p_T + \beta m}
\over {\sqrt{2} q^-}} \right\} \alpha^- u_\lambda (q)
\end{eqnarray}
where
\begin{equation}
\alpha^\pm = { 1 \over \sqrt{2}}(1 \pm \alpha_z) = \sqrt{2} P^\pm.
\end{equation}
Again $p^+ = (p_T^2 + m^2)/2q^-$
It is easy to see that this function solves the Dirac equation.
At $x^- = 0$, the term involving $p^+$ disappears so that the integrals
over $p_T$ and $z_T$ can be done with the result
\begin{equation}
	\psi(x) = {1 \over \sqrt{2}}
\left( \alpha^- + \alpha^+ {{\alpha_T \cdot Q_T + \beta m} \over {\sqrt{2}q^-}}
\right) e^{iqx} u_\lambda (q).
\end{equation}
Using the definition of the projection operators $P^\pm$, and the relationship
between $\psi\pm$, this is just $e^{iqx} u_\lambda (q)$.  
To derive this, we must use the definition of the vector potential in terms of
$\Lambda$.
This solution 
therefore solves the boundary conditions.

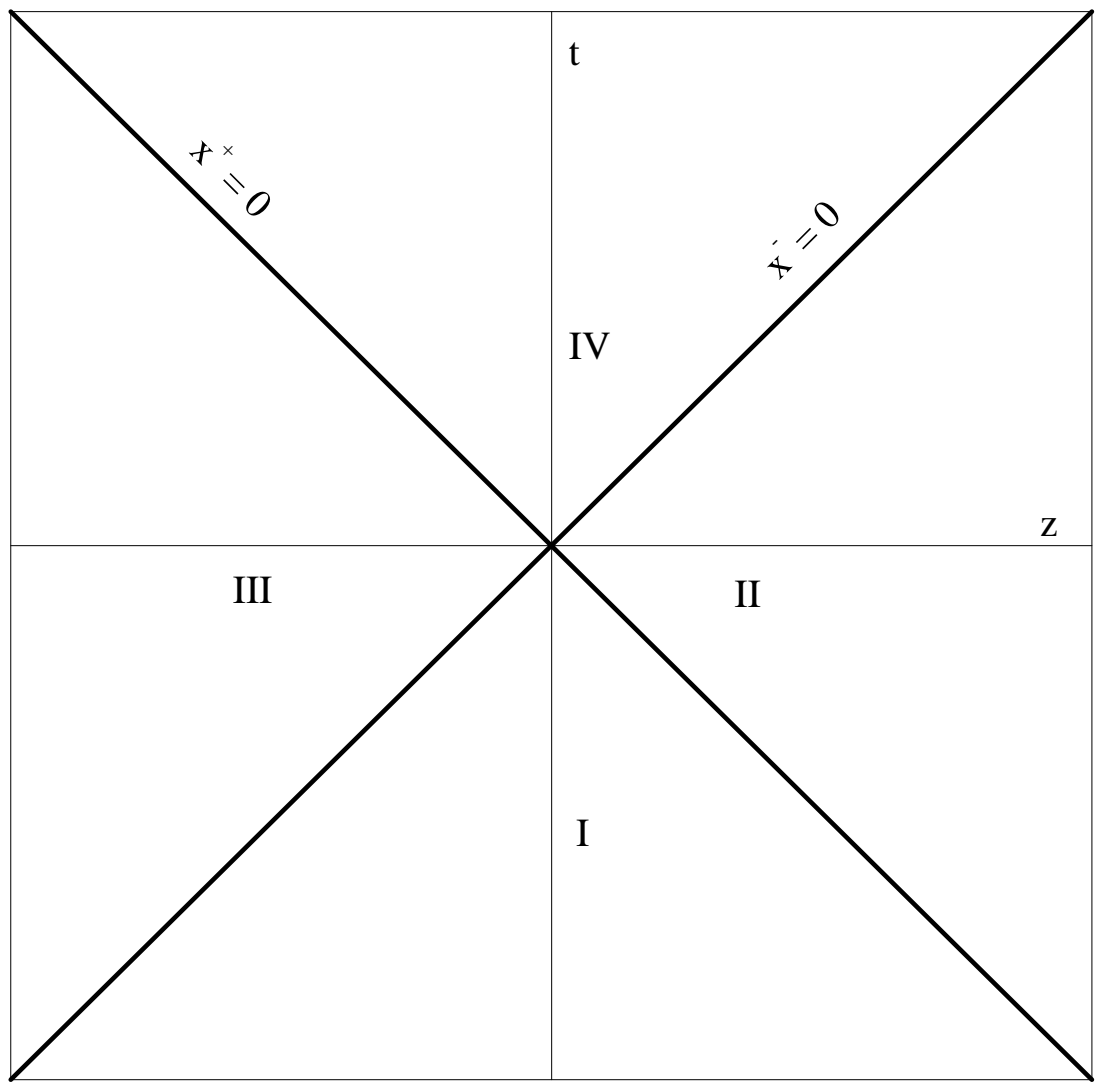
\begin{figure}
\caption[Figure 1]{Light cone boundaries of the four regions of the z, t plane
for an ultrarelativistic collision.}
\end{figure}
\end{document}